\documentclass{article}
\usepackage{amssymb}

\input{tcilatex}
\begin{document}

\title{Band structure and reflectance for a nonlinear one-dimensional
photonic crystal }
\author{S. Guti\'{e}rrez-L\'{o}pez, A. Castellanos-Moreno, A. Corella-Madue%
\~{n}o, \and and R. A. Rosas \\
Departamento de F\'{\i}sica,Universidad de Sonora, \\
Apartado Postal 1626, Hermosillo, Sonora, M\'{e}xico \and J. A. Reyes\thanks{%
On leave from Instituto de F\'{\i}sica, Universidad Nacional Aut\'{o}noma de
M\'{e}xico} \\
Departamento de F\'{\i}sica\\
Universidad Aut\'{o}noma Metropolitana,\\
Ixtapalapa, Apartado Postal 55 534 09340,\\
M\'{e}xico D. F., M\'{e}xico}
\date{}
\maketitle

\begin{abstract}
We consider a model for a one-dimensional photonic crystal formed by a
succession of Kerr-type equidistant spaceless interfaces immersed in a
linear medium. We calculate the band structure and reflectance of this
structure as a function of the incident wave intensity, and find two main
behaviors: the appearance of prohibited bands, and the separation and
narrowing of these bands. A system with these features is obtained by
alternating very thin slabs of a soft matter material with thicker solid
films, which can be used to design a device to control light propagation for
specific wavelength intervals and light intensities.

\end{abstract}

\newpage

Photonic crystals PCs ---spatially periodic, composite materials that can
exhibit photonic band PB gaps for light propagation---have reached a mature
state of development, with their optical properties well understood and with
many realized and potential applications \cite{1}. Most of this research
deals with PCs whose characteristics are fixed, that is, once they have been
fabricated there is no possibility to alter their optical response. A recent
trend, however, concerns tunable or active PCs; by this we imply that, by
means of some external agent, it becomes feasible to change the optical
properties of the PC continuously and reversibly. This could lead to tunable
optical waveguides, switches, limiters, and polarizers; to reconfigurable
optical networks; and to electrooptic interconnects in microelectronics. \
We can classify tunable PCs according to two broad categories. For one of
these, an external agent causes structural changes with no alteration of the
dielectric constants of the constituent materials. In the other category the
configuration of the PC remains the same, and it is some material property
of the PC that is affected by the external agent. Structural tuning has been
proposed or accomplished by means of mechanical stress applied to a polymer
opal of nanoscale spheres, by applying an electric field to a PC on a
piezoelectric substrate; by incorporating diodes in a 2D structure of wires;
and by applying a magnetic field to periodically distributed magnetic
particles \cite{dos}. On the other hand, tuning through the alteration of
some material property involves the incorporation of a ferromagnetic or
ferroelectric material in the PC -to be tuned by an external magnetic or
electric field, respectively \cite{tres}. In particular, LCs are
well-established electro-optic materials that can be tuned by means of
pressure, heat, and applied electric or magnetic field. Their incorporation
within a PC is of particular interest because of the possibility of
selectively tuning PB gaps, as reviewed recently \cite{3}.

In this letter we consider a system whose schematic plot is shown in
fig.1.This PC consists of a succession of spaceless nonlinear interfaces
immersed in a linear medium. The nonlinear interfaces are made of a soft
matter medium whose nonlinear response is that of a Kerr material for which
the nonlinear response is proportional to the square of the magnitude of the
intensity of the light \cite{Boyd}. A system with these features can be
managed by alternating very thin slabs of a soft matter material like a
nematic LC, with thicker solid films. Hence, for a low power laser beam (for
instance He-Ne) the nonlinear index refraction of the solid stacks is to be
negligible in comparison with that of the nematic films whose usual value is
more than eight orders of magnitude larger than that of a solid \cite{Khoo}.
In this model we shall ignore the thickness of the nonlinear slabs located
in $x=na$ and approximate their optical response by using the Kerr model.
This allow us to write the following propagation equation%
\begin{equation}
\frac{d^{2}\psi }{dx^{2}}+\left( k^{2}-\frac{n_{2}}{a}\sum_{n=-\infty
}^{\infty }\delta \left( x-na\right) \left\vert \psi \right\vert ^{2}\right)
\psi =0  \label{Schro}
\end{equation}%
where $n_{2}$ is the nonlinear refraction index and $\psi $ is the amplitude
of the propagating wave.\ For points $x\neq na$, the general solution of
this equation is of the form: $\psi =Ae^{ikx}+Be^{-ikx}$. However, after a
standard integration of Eq.(\ref{Schro}) around the points $x=na$ \cite%
{Gasiorowicz}, we can establish the boundary conditions%
\begin{equation}
\psi \left( 0\right) =e^{\chi a}\psi \left( a\right) ,e^{\chi a}\frac{d\psi
\left( a\right) }{dx}-e^{\chi a}\frac{d\psi \left( 0\right) }{dx}=e^{\chi
a}g\psi \left( 0\right) \left\vert \psi \left( 0\right) \right\vert ^{2},
\label{bc}
\end{equation}%
where we have already introduced the Bloch parameter \cite{Aschcroft} $\chi
a $ to characterize the momentum in each unit cell. Upon applying these
conditions to the general solution, permit us to write the following
transcendental equation 
\begin{equation}
f\left( z,p,\chi a\right) =p\frac{\cos \left( z-\chi a\right) \sin ^{3}z}{z}+%
\frac{1}{2}\cos \left( z-\chi a\right) \left[ 1-\cos \left( z+\chi a\right) %
\right] =0,  \label{tras}
\end{equation}%
where $z=ka$, $p=B^{2}gka$ and$\hspace{0.3cm}g=\frac{n_{2}}{2ka}$. \ In Fig.
2 we have illustrated the forbidden and permitted bands that we have found
by solving Eq.(\ref{tras}) numerically, where $f$ is plotted in the vertical
axis and $z$ is shown in the horizontal one.

Each plot is the result of the superposition of the different curves
obtained by varying the parameter $\chi a$ in the function $f\left( z,p,\chi
a\right) $. Hence, the roots of \ Eq.(\ref{tras}) correspond to the points
where curves cross the horizontal axis and thus each crossing give us a
point of the permitted band. Conversely, there is a prohibited band whenever
there is a white interval (no crossing point).

Fig. 3 depicts prohibited bands in the interval $0\leq z\leq 10$ and show
how they appear and disappear versus the parameter $p$.

Some interesting remarks of these plots are the following. Fig. 3a for the
interval $0\leq p\leq 1$ shows that there are no prohibited bands for $p=0$
whereas a prohibited band appears at $p=0.03$\ that widens,\ until a very
narrow prohibited band appears in $p=0.3$. For the interval $0.3\leq p<0.7$,
the first band is narrowed and the second one is widened. Simultaneously,
the first band is displaced to the left until the adjacent allowed band
located in $x=0$\ disappears.\ For $0.7\leq p\leq 0.9$,\ a third forbidden
band is added and it also widens against $p$. Fig. 3b for the interval $%
1\leq p\leq 1000$, \ a fourth prohibited band is added, and all of them
become narrower. For $p=100$ the left prohibited band is split into two
narrow bands, and for $p\geq 200$\ there are no more bands. In general terms
we can address two characteristic behaviors of these bands. one for the
appearance of the prohibited bands, and another for the separation and
narrowing of the mentioned bands.

Instead of an infinite array, for experimental purposes it is wealth to
consider a finite array of nonlinear interfaces $n=1,2,...$ immersed also in
a linear medium. It is convenient to write the general solution for a
nonlinear material in the following form%
\begin{equation}
\psi \left( x\right) =c_{n}e^{ikx}+d_{n}e^{-ikx},
\end{equation}%
where the subscript $n$ corresponds the stack number $n$. The boundary
conditions given by Eq.(\ref{bc}) can be expressed in the $n$-th interface as%
\begin{equation}
c_{n}e^{-ikx}+d_{n}e^{ikx}=c_{n-1}e^{-ikx}+d_{n-1}e^{ikx}  \label{c1}
\end{equation}%
and%
\begin{eqnarray}
&&-ikc_{n}e^{-ikx}+ikd_{n}e^{ikx}-\left(
-ikc_{n-1}e^{-ikx}+ikd_{n-1}e^{ikx}\right)  \nonumber \\
&=&g\left\vert c_{n-1}e^{-ikx}+d_{n-1}e^{ikx}\right\vert ^{2}\left(
c_{n-1}e^{-ikx}+d_{n-1}e^{ikx}\right) .  \label{c2}
\end{eqnarray}%
An additional condition stems from the fact that we shall assume that a wave
travels from the left, and no wave is coming from the right%
\begin{equation}
d_{0}=d_{n}\left( 0\right) =A,\hspace{0.3cm}c_{0}=c_{n}\left( 0\right) =B,%
\hspace{0.3cm}c_{n}\left( N\right) =0=C_{N}.  \label{ini}
\end{equation}

We solve the system of recurrence equations defined by Eqs.(\ref{c1}) and (%
\ref{c2}) with the initial conditions Eq.(\ref{ini}) by straightforward
substitution, and calculate directly the reflectance defined by $%
R=\left\vert B\right\vert ^{2}/\left\vert A\right\vert ^{2}$. Here we denote
by $n$ the number of non linear interfaces and $N-1$ the total number of
them contained in the finite array of stack. in this way the system will
have $N$ linear regions.

Fig. 4a shows the reflectance, $R\left( A,k\right) $ as function of the
independent variables $A$ and $k$. Fig. 4b exhibits the same information but
depicted in \ a contour plot. We present first the simplest case for which $%
n=1$ $\left( N=2\right) $, and $g=0.2$ to get some useful insight. It can be
observed that $R\left( A,k\right) $ goes to $1$, provided that $A$ tends to $%
4$,\ for any value of each $k$, except for the interval $2.1<k<2.3$, where
the material becomes transparent.

This region disappears when $g$ tends to $1$, as one can see in Fig. 5, to
reach finally the opacity ($R\left( A,k\right) \simeq 1$) for $g=6$, with $%
N=2$.

Fig. 6 displays a panel plots where $g$ and $N$ are varying. notice that by
keeping $g$ constant and increasing $N,$ the trough in \ the middle of the
surface is maintained.

Further calculations allow us to address the following interesting remarks.
For the case of a only one nonlinear interface $\left( N=2\right) $, $%
R\left( A,k\right) $ increases by enlarging $A$ and keeping constant the
value of $k.$ Nevertheless, there exists a transparency stripe situated
approximately in the $k-$interval $[2,3]$ which narrows when $g$ grows until
it disappears for $g=6$. Similarly, for the case of three nonlinear
interfaces $\left( N=4\right) $: $R\left( A,k\right) $ grows for $k$ fixed
by enlarging $A$. Also, there is a transparency stripe which is narrowing \
but it does not disappear even for $g=6$. The behavior is similar even when
the system consists of five nonlinear interface $\left( N=6\right) $. A
general feature observed in all these calculations is that by increasing
both $g$ and $N$, the border of the transparency band is no longer smooth.

We have elaborated a model for a simple nonlinear one-dimensional PC for
which the nonlinearity is concentrated in certain spaceless barriers
periodically located. This system can be constructed by piling up slabs of a
nonlinear soft matter material like a LC inserted between thicker solid
films. We have calculated analytically the band structure of this periodic
structure. We have shown that by increasing the intensity of the normal
incident wave, there appear more prohibited bands which separate each other
and get narrow. The same qualitative behavior is shown by the reflectance of
a finite array of alternating stacks for which forbidden bands displace and
narrow versus the signal intensity.

Our results suggest to design devices based on this optical structure in
order to prevent the propagation of light whose intensity is either larger
or lower than certain threshold, for a specific given interval of the
wavelength spectrum.

\begin{description}
\item Fig. 1 Schematic plot of the nonlinear photonic crystal. The
interfaces exhibit Kerr-type optical response whereas the rest of the
material displays a linear response.

\item Fig. 2 Band structure of the nonlinear PC. $f\left( z,p,\chi a\right) $
versus $z$ parametrized by $\chi a$. The roots of Eq.(\ref{tras}) correspond
to the points where curves cut the $z-$axis and thus each crossing give us a
point of the permitted band. a) $p=0.03$, b) $p=0.3$, c) $p=0.7$ and d) $%
p=0.9$.

\item Fig.3 The clearer intervals denote the forbidden bands versus light
intensity $p$ (vertical axis) in the interval $0\leq z\leq 10$. a) $0\leq
p\leq 1$ and b) $0\leq p\leq 1000$.

\item Fig 4 a) Reflectance $R$ against $A$ and $k$ $\ $for $n=1$ $\left(
N=2\right) $, and $g=0.2$. b) Contour plot of $R$ versus $A$ and $k$ and the
same parameters. The clearer regions correspond to larger values of $R$.

\item Fig 5 The same as Fig. 4 but for $g=1$ and $n=1(N=2)$.

\item Fig. 6 Table of contour plots of $R$ against $A$ and $k$ for the shown
values of $g$ and $N$.
\end{description}

\end{document}